\documentclass[a4paper,11pt]{article}
\usepackage{jcappub}

\renewcommand{\thesection}{\arabic{section}}

\usepackage{hyperref}
\usepackage{amsmath}
\usepackage{amssymb}
\usepackage{graphicx}
\usepackage{color}
\usepackage{subfig}
\usepackage{bm}
\usepackage{gensymb}
\usepackage{bigints}
\usepackage[usenames,dvipsnames,svgnames,table]{xcolor}
\usepackage{pifont}
\usepackage{aas_macros}
\usepackage{orcidlink}
\usepackage{physics}
\usepackage{makecell}
\usepackage{booktabs}

\usepackage[labelfont=bf]{caption}

\hypersetup{
    colorlinks=true,
    linkcolor=Blue,
    filecolor=Blue,
    urlcolor=MidnightBlue,
    citecolor=Blue,
   pdftitle={Cooling age}
}
\renewcommand{\thetable}{\arabic{table}}

\arxivnumber{}

\begin{document}


\title{Cosmological constraints of Palatini $f(\mathcal{R})$ gravity}


\author[a]{D{\'e}bora Aguiar Gomes,}
\author[b]{Rebecca Briffa,}
\author[c]{Aleksander Kozak,}
\author[b]{Jackson Levi Said,}
\author[a]{Margus Saal}
\author[d]{and Aneta Wojnar\orcidlink{0000-0002-1545-1483}}

\affiliation[a]{Laboratory of Theoretical Physics, Institute of Physics, University of Tartu,\\
W. Ostwaldi 1, 50411 Tartu, Estonia}
\affiliation[b]{Institute of Space Sciences and Astronomy, University of Malta, Malta, MSD 2080}
\affiliation{Department of Physics, University of Malta, Malta}
\affiliation[c]{Institute of Theoretical physics, University of Wroclaw, pl. Maxa Borna 9, 50-206 Wroclaw, Poland}
\affiliation[d]{Departamento de F\'isica Te\'orica \& IPARCOS, Universidad Complutense de Madrid, E-28040, 
Madrid, Spain}

\emailAdd{debora.aguiar.gomes@ut.ee}
\emailAdd{rebecca.briffa.16@um.edu.mt}
\emailAdd{aleksander.kozak@uwr.edu.pl}
\emailAdd{jackson.said@um.edu.mt}
\emailAdd{margus.saal@ut.ee}
\emailAdd{awojnar@ucm.es}

\abstract{In this study, we investigate a Palatini $f(R)$ gravity model featuring a quadratic term correction, aligning it with the most recent expansion rate data, with a particular focus on the latest SNIa and BAO data. Our analysis employs CC data as the fundamental dataset, complemented by contributions from the SN sample and a combination of non-overlapping transversal BAO datasets. We conduct a comprehensive MCMC analysis for each data set combination, yielding constraints on all cosmological parameters within the model. Additionally, we incorporate the latest Hubble constant value from the SH0ES Team. Finally, we present a statistical comparison between the Palatini quadratic model and $\Lambda$CDM using the AIC and BIC metrics, ultimately obtaining {the constraint $|\alpha| \leq 10^{49}\,\text{m}^2$}. We also stress the significance of studying stellar and substellar objects for obtaining more precise constraints on modified gravity compared to those derived from cosmological observations.}

\maketitle

\section{Introduction}

The standard $\Lambda$CDM concordance model continues to be the strongest model that is consistent with observational measurements at all scales of cosmology \cite{Misner:1974qy,Clifton:2011jh}. In this setting, cold dark matter (CDM) acts on large scale structures \cite{Peebles:2002gy,Baudis:2016qwx,Bertone:2004pz}, whereas dark energy takes on the form of a cosmological constant ($\Lambda$) \cite{Copeland:2006wr}, and acts in the late Universe to produce cosmic acceleration. While great efforts have been put into the cosmological constant, internal consistency issues persist \cite{Weinberg:1988cp}. On the other hand, dark matter particles remain elusive in particle physics observatories \cite{Gaitskell:2004gd}. In addition, other important problems plague the $\Lambda$CDM model such as the lithium problem \cite{DiBari:2013dna} in Big Bang nucleosynthesis measurements, as well as anomalies in the cosmic microwave background (CMB) radiation ranging from the lack of large-scale temperature correlations \cite{doi:10.1073/pnas.90.11.4766} to asymmetries in all-sky surveys \cite{2020ApJS..247...69E, Eriksen:2007pc}, as well as other open questions \cite{Aghanim:2018eyx,Perivolaropoulos:2021jda}. Besides these long-standing issues, recently, the effectiveness in the ability of $\Lambda$CDM to describe certain combinations of data sets has been called into question \cite{DiValentino:2020vhf,DiValentino:2020zio,DiValentino:2020vvd,DiValentino:2020srs}. This is called the cosmic tensions problem, and poses a foundational challenge to the $\Lambda$CDM concordance model. The tension between direct measurements of cosmic expansion in the late Universe \cite{Riess:2019cxk,Anderson:2023aga,Wong:2019kwg}, and model-dependent observations using data from the early Universe \cite{Aghanim:2018eyx,ACT:2023kun,Schoneberg:2022ggi} is a growing statistical feature of recent studies. While it may be that this is an artefact of the type of measurement technique being employed \cite{Riess:2019qba,Pesce:2020xfe,deJaeger:2020zpb,Capozziello:2023ewq}, but the consistency of the appearance of this tension seems to indicate the potential requirement for a new physical model of cosmology \cite{Abdalla:2022yfr,Bernal:2016gxb}.

The combined challenges to the $\Lambda$CDM model has led to a re-evaluation of competing formulations of different cosmological models \cite{Sotiriou:2008rp,Clifton:2011jh,CANTATA:2021ktz,Krishnan:2020vaf,Colgain:2022nlb,Malekjani:2023dky,Ren:2022aeo,Dainotti:2021pqg,Addazi:2021xuf,Schoneberg:2021qvd}. The paradigm of early- and late-Universe emerges out of the cosmic tensions problem itself, and has prompted attempts at resolutions to the problem that focus in these separate sectors individually. Late-time solutions alter the evolution of the Universe while preserving its early-time behavior. Models of this nature range from decaying dark matter models \cite{Anchordoqui:2020djl}  as well as decaying dark energy models \cite{Alam:2003rw}. These sometimes include interaction terms \cite{Gariazzo:2021qtg, Piedipalumbo:2023dzg}. There are also running vacua models \cite{Sola:2016jky}, as well as cosmological constant sign switching models \cite{Akarsu:2023mfb}. However, the combination of supernovae of Type Ia (SNIa), cosmic chronometer (CC), and baryonic acoustic oscillation (BAO) data severely curtails the possibility of resolving the cosmic tensions problem using only late-time modifications to cosmology \cite{Colgain:2023bge}.

On the other end of the spectrum there have been many early Universe proposals which are principally focused on decreasing the sound horizon, since this infers a higher Hubble constant for a fixed CMB angular scale. Early Universe dark energy models \cite{Poulin:2023lkg} are the most studied proposals in this sector, whereas models of neutrinos beyond the standard particle physics model are also potentially active in this field \cite{DiValentino:2017oaw}, among many others. The decrease in the sound horizon does pose a problem for these models since it may jeopardize the compatibility of these cosmological models with the growth of large scale structures \cite{Jedamzik:2020zmd}. Altogether, this points to a global rescaling of the Hubble evolution of the Universe \cite{Escamilla-Rivera:2019hqt}. This may include a combination of modifications to the gravitational and matter sectors.

In recent years, there have been several proposals for a diversity of different classes of modifications to the gravitational sector, which have been shown to tackle a range of physical problems \cite{CANTATA:2021ktz,Capozziello:2002rd,Capozziello:2011et,Clifton:2011jh,Nojiri:2010wj,Nojiri:2017ncd}. One interesting class of models are Ricci-based ones, among which Palatini $f(R)$ is the simplest, apart from General Relativity (GR), and the most studied extensions of the Einstein's proposal.

Research in Palatini $f(R)$ gravity, which is comprehensively reviewed in 
\cite{Olmo:2011sw}, has encompassed a wide spectrum of cosmological inquiries. Notably, \cite{Olmo:2008nf} unveiled the derivation of the Palatini effective Lagrangian from Loop Quantum Cosmology, shedding light on its covariant characteristics. The authors of \cite{Szydlowski:2015fcq} delved into the examination of singularities and constraints, particularly with respect to the possibility of inflation, employing Chaplygin gas models. The exploration of bouncing solutions and perturbations within this framework was documented in \cite{Koivisto:2010jj}. In contrast, \cite{Capozziello:2018aba} focused on the reconstruction of a class of Palatini Lagrangians using cosmography and analytical approximations, incorporating Planck data to determine key parameters such as $\Omega_m$. The consideration of torsion, as illustrated in \cite{Bombacigno:2018tyw}, extended the theory, leading to scalar-tensor Palatini gravity, though it raised concerns about its stability. The examination of Noether symmetries in a cosmological context was the subject of \cite{Roshan:2008tt}. Early-time inflation and late-time cosmic acceleration without invoking dark energy were investigated in \cite{Sotiriou:2005hu}, while the derivation of the effective equation of state in Palatini cosmology was presented in \cite{Camera:2022myt}.

Furthermore, an avenue of research, as exemplified by the studies \cite{Barragan:2009sq,Barragan:2010qb}, probed into bouncing cosmologies, scrutinizing scenarios both isotropic and anisotropic. Simultaneously, \cite{Koivisto:2007sq} centered its focus on comprehending the formation of large-scale structures. Additionally, research conducted by a separate group \cite{Uddin:2007gj} was dedicated to exploring cosmological perturbations within the Palatini $f(R)$ gravity framework. Lastly, a distinct approach to the study of cosmological singularities was undertaken by another set of researchers \cite{Szydlowski:2017rjx}, significantly enhancing our understanding of the singular behavior inherent in this gravitational model.

On the other hand, research in Palatini gravity with scalar fields has also yielded a diverse array of insights. Among others, \cite{deSouza:2013uu} explored the non-minimal coupling of scalar fields in Palatini gravity, employing a Noether symmetry approach.
Additionally, the concept of Palatini Higgs inflation has been rigorously examined in a series of studies \cite{Bauer:2010jg,Rasanen:2017ivk,Tenkanen:2019jiq,Shaposhnikov:2020fdv,Gialamas:2020vto}.
Furthermore, the interplay of inflationary dynamics and Palatini quadratic gravity has been extensively explored in various works, including \cite{Meng:2004yf,Enckell:2018hmo,Antoniadis:2018ywb,Bostan:2019uvv}, and others. 
The specific dynamics of tachyonic preheating in Palatini $R^2$ inflation were elucidated in \cite{Karam:2021sno}.
Moreover, quintessential inflation within Palatini $f(R)$ gravity was analyzed in \cite{Verner:2020gfa,Dimopoulos:2020pas,Dimopoulos:2022rdp}.
Lastly, the comparison between metric and Palatini formulations of inflation was studied in various works, such as \cite{Bauer:2008zj,Jarv:2020qqm,Karam:2021wzz,Bostan:2022swq,Eadkhong:2023ozb}.

The main purpose of this paper is to test the well-known Starobinsky (quadratic) model in the Palatini approach in the context of late-time cosmic expansion by comparing its predictions against empirical data and performing a suitable statistical analysis. The next section introduces the necessary theoretical background, including the aforementioned Palatini approach to modified gravity, that will elucidate the key differences between the model considered here and the $\Lambda$CDM model. We present the modified Friedmann equation, that will be the basis for our numerical calculations. Next, the data sets used in our investigations will be described, as well as the methodology allowing us to impose constraints on the parameters introduced by the model. We then employ an MCMC analysis of the quadratic model and present the results for the Hubble parameter, matter content, the absolute magnitude of SNIa, and the Starobinsky parameter $\alpha$, accompanied by two statistical criteria assessing the performance of the model compared to $\Lambda$CDM. At the end, we draw some conclusions and summarize the results of the paper.

\section{Palatini gravity}\label{s1}

\subsection{Preliminaries}

Palatini gravity belongs to the category of metric-affine models of gravity, representing the simplest example of this framework. In this approach, the metric tensor $g$ and the connection $\hat\Gamma$ are considered as independent entities. Consequently, to derive the field equations governing this gravitational theory, one must perform variations with respect to both of these variables. The action functional for this theory is given as follows:
\begin{equation}\label{Eq:f(R)}
S = \frac{1}{2 \kappa^2} \int d^4x \sqrt{-g} f(\mathcal{R}) + S_m[g_{\mu\nu},\psi_m].
\end{equation}
Here, $\sqrt{-g}$ represents the determinant of the metric tensor, $f(\mathcal{R})$ is a function of the curvature scalar $\mathcal{R}$, and $S_m$ represents the action for matter fields $\psi_m$, which exclusively depends on the metric tensor $g_{\mu\nu}$. Importantly, the matter action $S_m$ remains entirely independent of the connection $\hat\Gamma$. Although it is conceivable to extend the matter action to encompass a dependence on $\hat\Gamma$, such an augmentation is deemed pertinent primarily in the context of fermionic interactions as opposed to classical fluid dynamics on a large scale, and thus, we shall omit its consideration in this context.

Furthermore, the curvature scalar $\mathcal{R}$ is constructed utilizing both the metric tensor and the Palatini-Ricci curvature tensor $\mathcal{R}_{\mu\nu}(\Gamma)$. Specifically, it can be expressed as $\mathcal{R} = g^{\mu\nu}\mathcal{R}_{\mu\nu}(\Gamma)$. It's noteworthy that $\mathcal{R}_{\mu\nu}$ must maintain symmetry to prevent instabilities \cite{Afonso:2017bxr, BeltranJimenez:2019acz, BeltranJimenez:2020sqf}.

Should the functional form of $f$ exhibit linearity with respect to $\mathcal{R}$, we recover GR. In scenarios involving vacuum or spacetime dominated by pure radiation, we effectively deal with GR featuring a cosmological constant. However, when considering $f$ as an arbitrary function of $\mathcal{R}$, we encounter a distinct spacetime structure \cite{Allemandi:2004ca, Allemandi:2004wn, Olmo:2011sw}. Significantly, within our context, these modifications carry implications for the description of stellar phenomena \cite{Olmo:2019flu} and early and late cosmology \cite{Allemandi:2004ca,Allemandi:2004wn,Allemandi:2005qs,Borowiec:2011wd}. Let us delve deeper into this subject.

Variation of Eq. \eqref{Eq:f(R)} solely with respect to the metric tensor $g_{\mu\nu}$ results in the following field equations:
\begin{equation}
f'(\mathcal{R})\mathcal{R}_{\mu\nu}-\frac{1}{2}f(\mathcal{R})g_{\mu\nu}=\kappa^2 T_{\mu\nu},\label{structural}
\end{equation}
where $T_{\mu\nu}$ represents the energy-momentum tensor for the matter field and can be derived in the conventional manner:
\begin{equation}\label{Eq: energytensor}
T_{\mu\nu}=-\frac{2}{\sqrt{-g}}\frac{\delta S_m}{\delta g_{\mu\nu}}.
\end{equation}
In our scenario, we assume the matter content to be in the form of a perfect fluid. In the aforementioned equations, the primes denote derivatives with respect to the argument of the function, denoted as $f'(\mathcal{R})=df(\mathcal{R})/d\mathcal{R}$. Evidently, for linear $f(\mathcal{R})$, these equations reduce to the familiar equations of GR.

Conversely, the variation with respect to the independent connection $\hat\Gamma$ can be expressed as \cite{Allemandi:2005qs,Wojnar:2017tmy}:
\begin{equation}
\hat\nabla_\beta(\sqrt{-g}f'(\mathcal{R})g^{\mu\nu})=0.\label{con}
\end{equation}
Here, $\hat\nabla_\beta$ denotes the covariant derivative calculated with respect to $\hat\Gamma$. Notably, by introducing a new metric tensor defined as:
\begin{equation}\label{met}
\hat{g}_{\mu\nu}=f'(\mathcal{R})g_{\mu\nu},
\end{equation}
we can reformulate Eq. \eqref{con} as follows:
\begin{equation}\label{met2}
\nabla_\beta(\sqrt{- \hat{g}} \hat{g}^{\mu\nu})=0,
\end{equation}
provided that the independent connection $\hat\Gamma$ corresponds to the Levi-Civita connection associated with the metric $\hat g_{\mu\nu}$. It is also evident that, for linear $f$, the connection $\hat\Gamma$ reduces to the Levi-Civita connection $\Gamma$ associated with the metric $g_{\mu\nu}$.

Furthermore, we can gain deeper insights into the characteristics of this gravitational theory by scrutinizing the $g$-trace of Eq. \eqref{structural}:
\begin{equation}
f'(\mathcal{R})\mathcal{R}-2 f(\mathcal{R})=\kappa^2 T,\label{struc}
\end{equation}
where $T$ represents the trace of the energy-momentum tensor. This expression unveils the possibility of algebraically expressing $\mathcal{R}$ as a function of matter fields, contingent on a given $f(\mathcal{R})$. In instances featuring vacuum and/or pure radiation-dominated spacetime, where $T=0$, we can solve Eq. \eqref{struc}, resulting in the theory converging to GR with a cosmological constant. Additionally, it can be established that this theory does not introduce any supplementary degrees of freedom beyond those already inherent in GR. This contrasts with the case of metric $f(R)$ gravity, where the relationship between $R$ and $T$ arises through a differential equation, thereby introducing an extra degree of freedom, often interpreted as a scalar field, subject to appropriate transformations.

In our subsequent analysis, we will consider the simplest extension of GR, often referred to as the Starobinsky or quadratic model, which is expressed as:
\begin{equation} \label{Eq:fR Quadratic}
f(\mathcal R) = \mathcal R + \alpha \mathcal R^2,
\end{equation}
where $\alpha$ serves as a parameter. In this scenario, the solution to Eq. \eqref{struc} adopts the form of GR, i.e., $\mathcal R = -\kappa^2 T$. This signifies that any modifications manifesting in the field equations are contingent on matter fields and are parametrized by a solitary parameter, $\alpha$.

\subsection{Palatini cosmology} \label{sec:palatini_cosmo}

As was already mentioned, Palatini $f(\mathcal{R})$ gravity does not introduce any additional degrees of freedom and reduces to GR with a cosmological constant in the vacuum. Due to this fact, Palatini theories find applications in modeling the late-time acceleration of the Universe \cite{Olmo:2011sw, Gogoi:2021mhi}, where curvature functions of type $f(\mathcal{R}) = \mathcal{R} + \alpha_n \mathcal{R}^n$ are of particular interest. It was also found that the choice $f(\mathcal{R}) = \mathcal{R}^{n(t)}$, where the exponent depends on the temperature and varies from $n=1$ in the low-energy limit, and $n=2$ at the early Universe, yields a scale-invariant description of gravity in high-energy limit, reduces to GR at later times, and evades Ostroradsky's instability \cite{Coumbe:2019fht}.


In what follows, we will assume the quadratic model in the Palatini approach, i.e. $f(\mathcal{R}) = \mathcal{R} + \alpha \mathcal{R}^2$, and the standard Friedmann-Lema\^{i}tre-Robertson-Walker metric:
\begin{equation}
    g_{\mu\nu} = \text{diag}\left(-1, \frac{a^2}{1-kr^2}, a^2 r^2, a^2 r^2 \sin^2 \theta\right)
\end{equation}
and the stress-energy tensor for the perfect fluid:
\begin{equation}
    T_{\mu\nu} = (p + \rho)u_\mu u_\nu + p g_{\mu\nu}, \quad T = 3p - \rho,
\end{equation}
satisfying the continuity equation $\nabla^{(g)}_\mu T^{\mu\nu} = 0$, where $\nabla^{(g)}$ is the covariant derivative calculated using the metric tensor $g_{\mu\nu}$. The continuity equation allows us to write:
\begin{equation}\label{rpal}
    \mathcal{R}=3H_0^2(\Omega_{m,0} a^{-3} + 4\Omega_{\Lambda, 0}) \,.
\end{equation}
It must be noted that the two Ricci tensors, Palatini $\mathcal{R}_{\mu\nu}$ and metric $R_{\mu\nu}$, are related to each other via a conformal transformation since $\mathcal{R}_{\mu\nu} = \mathcal{R}_{\mu\nu}(f'(T) g)$, so that\footnote{Let us notice that the curvature scalar $\mathcal{R}$ is not the curvature scalar built from the conformal metric only, $\hat{R}(f'(T) g)$, but rather it is a contraction of $g^{\mu\nu}$ and the Palatini Ricci tensor $\mathcal{R}_{\mu\nu}(f'(T) g)$.} \cite{Sotiriou2010}
\begin{equation}\label{rmunu}
    \begin{split}
        \mathcal{R}_{\mu\nu} & = R_{\mu\nu} + \frac{3}{2}\frac{1}{(f'(T))^2}\partial_\mu f'(T)\partial_\nu f'(T) \\
        &- \frac{1}{f'(T)}\left(\nabla^{(g)}_\mu \nabla^{(g)}_\nu  - \frac{1}{2}g_{\mu\nu}\Box^{(g)} \right) f'(T) \,.
    \end{split}
\end{equation}
To get the Friedmann equation, one needs to plug relations \eqref{rpal} and \eqref{rmunu} to the field equations \eqref{structural} and get rid of $a''$ from one of the formulas. Upon doing so, one arrives at the following equation for the quadratic model \cite{Borowiec:2011wd, Borowiec:2015qrp,Szydlowski:2015fcq,Stachowski:2016zio}:
\begin{align}
\frac{H^2}{H_0^2}&=\frac{b^2}{\left(b+\frac{d}{2}\right)^2}\left[\Omega_{\alpha}(\Omega_{\text{m},0}a^{-3}+\Omega_{\Lambda,0})^2 
\times \frac{(K-3)(K+1)}{2b} \right.\nonumber \\ 
&\left. +(\Omega_\text{m,0}a^{-3}+\Omega_{\Lambda,0}) 
 +\frac{\Omega_{\text{r},0}a^{-4}}{b}+
\Omega_k\right],\label{friedmann2}
\end{align}
where
\begin{align}
\Omega_k &= -\frac{k}{H_0^2 a^2},\\
\Omega_{\text{r},0} &= \frac{\kappa^2\rho_\text{r,0}}{3H_0^2},\\
\Omega_{\text{m},0} &= \frac{\kappa^2\rho_\text{m,0}}{3H_0^2},\\
\Omega_{\Lambda,0} &= \frac{\kappa^2\rho_{\Lambda,0}}{3H_0^2},\\
K &= \frac{3\Omega_{\Lambda,0}}{(\Omega_\text{m,0}a^{-3}+\Omega_{\Lambda,0})},\\
\Omega_{\alpha} &= 3\alpha H_0^2,\\
b &= f'(\hat{R})=1+2\Omega_\alpha(\Omega_\text{m,0}a^{-3}+4\Omega_{\Lambda,0}),\\
d &= \frac{1}{H}\frac{db}{dt}=-2\Omega_{\alpha}(\Omega_\text{m,0}a^{-3}+\Omega_{\Lambda,0})(3-K) \,.
\end{align}

From the above one can check that the standard model (\ref{lcdm}) can be reconstructed in the limit $\alpha\mapsto 0$:
\begin{equation}\label{lcdm}
\frac{H^2}{H_0^2}=\Omega_{\text{r},0}a^{-4}+\Omega_{\text{m},0}a^{-3}+\Omega_{\Lambda,0}+\Omega_{k}.
\end{equation}
In what follows, we will skip the spatial curvature term $\Omega_k$.

\section{Constraints from Observational Data}

In this section, we start with a description of the observational data that we use together with a brief explanation of how we implement this into our analysis pipeline. We then go on to the main constraint results which is composed of a series of Markov Chain Monte Carlo (MCMC) constraint analyses which we use to both set limits on the parameters in the model, with an emphasis on the new $\Omega_{\alpha}$ parameter, as well as to assess how the model compared with the standard $\Lambda$CDM cosmological model in each case.

\subsection{Observational Data}

We devote this part of the manuscript to the description of the data sets being used in our analysis. Our baseline data set consists of Hubble expansion data which we then complement with supernova Type Ia (SNIa) and Baryonic acoustic oscillation (BAO) data. We also include the recent release of the SH0ES value for the Hubble constant \cite{Riess:2021jrx}, $H_0^{\rm S22} = 73.04 \pm 1.04 \,{\rm km\, s}^{-1} {\rm Mpc}^{-1}$, which has an associated reported absolute magnitude of $M^{\rm S22} = -19.253 \pm 0.027 \,{\rm mag}$. The other data sets are
\begin{itemize}
    \item \textbf{Cosmic Chronometers (CC)} -- We use the public list of $31$ data points \cite{2014RAA....14.1221Z,Jimenez:2003iv,Moresco:2016mzx,Simon:2004tf,2012JCAP...08..006M, Stern:2009ep,Moresco:2015cya} that utilize the CC method. In this approach, a spectroscopic dating technique is used for passively evolving galaxies. In this way, direct measurements are obtainable for the Hubble parameter up to $z \lesssim2$. CC also has the advantage that it is independent of any cosmological model as well as the the Cepheid distance scale. On the other hand, some assumptions are made on the modeling of stellar ages. For two passively aging galaxies, observations can constrain $\Delta z/ \Delta t$ which can inturn be used to estimate the Hubble parameter at that redshift through $H(z) = -(1+z)^{-1} \Delta z/ \Delta t$. The corresponding $\chi^2_H$ estimator is represented by
    \begin{equation}
        \chi^2_H = \sum^{31}_{i=1} \frac{\left(H(z_i,\Theta) - H_{\mathrm{obs}}(z_i)\right)^2}{\sigma^2_H(z_i)} \,,
    \end{equation}
    where $H(z_i, \Theta)$ are the theoretical model Hubble parameter values at redshift $z_i$ with model parameters $\Theta$ while $H_{\mathrm{obs}}(z_i)$ are the corresponding Hubble data values at $z_i$ with an associated uncertainty of $\sigma_H(z_i)$. 
    
    \item \textbf{Pantheon+ Compilation} -- This compilation is based on SNIa observations that are calibrated using a Cepheid distance scale. These events occur in binary stellar systems with uniform intrinsic brightness making them ideal standard candles in order to measure expansion profilers for distant galaxies. The Pantheon+ (SN) compilation \cite{Scolnic:2021amr} consists of 1701 SNIa samples which reports distance modulus $\mu$ of an object together with the redshift $z$ as measured in the frame of the cosmic microwave background radiation (CMB). The distance modulus depends on the observed apparent magnitude of an object, $m$, together with its absolute magnitude, $M$ which is a measure of intrinsic brightness. For an object at redshift $z_i$, the distance modulus is given as
    \begin{equation}
        \mu(z_i, \Theta) = m - M = 5 \log_{10}[D_L(z_i, \Theta)] + 25 \,,
    \end{equation}
    where $D_L(z_i, \Theta)$ is the luminosity distance defined as 
    \begin{equation}
        D_L(z_i, \Theta) = c(1+z_i) \int_0^{z_i} \frac{dz'}{H(z', \Theta)} \,. 
    \end{equation}
    The SN data set needs to be calibrated for the absolute magnitude $M$, which is treated as a nuisance parameter in the MCMC and is marginalized over. Finally, the cosmological parameters are constrained by minimizing a $\chi^2$ likelihood specified by \cite{SNLS:2011lii}
    \begin{equation}
        \chi^2_{\mathrm{SN}} = (\Delta \mu(z_i), \Theta))^T C^{-1} (\Delta \mu(z_i), \Theta)) \,,
    \end{equation}
    where $(\Delta \mu(z_i), \Theta)) = (\mu(z_i), \Theta) - \mu(z_i)_{\mathrm{obs}}$ and $C$ is the corresponding covariance matrix which accounts for the statistical and systematic uncertainties. The SN compilation used in this analysis incorporates the SH0ES Cepheid host distance anchors ($H_0^{\rm S22}$) \cite{Riess:2021jrx} in the likelihood which helps to break the degeneracy between the parameters $M$ and $H_0$ when analyzing SNIa alone. The redshift range of the SN compilation is $0.01 < z < 2.5$ which includes much more information at lower redshift as compared with previous SNIa compilations.
    
    \item \textbf{Baryon Acoustic Oscillations} - Our analysis is nuanced by a joint BAO data set consisting of separately independent data points. This includes measurements from the  SDSS Main Galaxy Sample at $z_{\mathrm{eff}} = 0.15$ \cite{Ross:2014qpa}, the six-degree Field Galaxy Survey at $z_{\mathrm{eff}} = 0.106$ \cite{2011MNRAS.416.3017B}, and the BOSS DR11 quasar Lyman-alpha measurement at $z_{\mathrm{eff}} = 2.36$ together with the BOSS DR10 measurements at $z_{\mathrm{eff}} = \{0.32,0.57\}$ \cite{BOSS:2013rlg}. We also incorporate the angular diameter distances and $H(z)$ measurements of the SDSS-IV eBOSS DR14 quasar survey at $z_{\mathrm{eff}} = \{0.98, 1.23, 1.52, 1.94\}$ \cite{Zhao:2018gvb}, along with the SDSS-III BOSS DR12 consensus BAO measurements of the Hubble parameter and the corresponding comoving angular diameter distances at $z_{\mathrm{eff}} = \{0.38, 0.51, 0.61\}$ \cite{Alam:2016hwk}. These two BAO data sets include covariance matrix information which is included in our MCMC analysis.
    The BAO data sets involve various measures of cosmic expansion include the Hubble distance $D_H(z)$, comoving angular diameter distance $D_M(z)$, and volume-average distance $D_V(z)$, which are respectively defined through
    \begin{align}
        D_H(z) &= \frac{c}{H(z)}\,, \nonumber\\
        D_M(z) &= (1+z)D_A(z)\,, \nonumber\\
        D_V(z) &= \left[(1+z)^2D_A(z)^2 \frac{c z}{H(z)}\right]^{1/3} \,,
    \end{align}
    and where $D_A(z)=(1+z)^{-2}D_L(z)$ is the angular diameter distance. In accordance with the BAO results, we calculate the combination of parameters
    $\mathcal{G}(z_i)=D_V(z_i)/r_s(z_d)$,
    $r_s(z_d)/D_V(z_i),$
    $D_H(z_i),$ 
    $D_M(z_i)(r_{s,\mathrm{fid}}(z_d)/r_s(z_d)),$
    $H(z_i)(r_s(z_d)/r_{s,\mathrm{fid}}(z_d)),$ \\ 
    $D_A(z_i)(r_{s,\mathrm{fid}}(z_d)/r_s(z_d))$ 
    for which the comoving sound horizon at the end of the baryon drag epoch, at redshift $z_d\approx 1059.94$, has a fiducial value of $z_d\approx 1059.94$ \cite{Planck:2018vyg}. This lets us define the corresponding $\chi^2$ for the BAO data as
    \begin{equation}
        \chi^2_{\text{BAO}}(\Theta) = \Delta G(z_i,\Theta)^T C_{\text{BAO}}^{-1}\Delta G(z_i,\Theta)
    \end{equation}
    where $\Delta G(z_i,\Theta) = G(z_i,\Theta)-G_{\text{obs}}(z_i)$and $C_{\text{BAO}}$ is the covariance matrix of all the considered BAO observations.
\end{itemize}

Finally, we include statistical criterions through which to assess the performance of the model against $\Lambda$CDM for the various data sets as well as the inclusion of the $H_0^{\rm S22}$ prior. To do this, we use the Akaike Information Criteria (AIC) \cite{1100705} which includes the best performing minimum $\chi^2_\mathrm{min}$ values and a measure of the model complexity, through the number of parameters $n$. This goodness-of-fit measure is then defined as 
\begin{equation}
    \mathrm{AIC} = \chi^2_\mathrm{min} + 2n \,.    
\end{equation}
AIC penalizes high numbers of parameters and so lower values of AIC are indicative of better performing models.

We also examine the Bayesian Information Criterion (BIC) which nuances the AIC by putting more emphasise on the level of complexity that a model exhibits. BIC is defined as
\begin{equation}
    \mathrm{BIC} = \chi^2_\mathrm{min} + n \ln m \,,
\end{equation}
where $m$ is the sample size of the observational data combination. The goals of both parameters is the same but BIC helps penalize higher parameters models in a stronger way. Absolute values of these statistical measures are difficult to relate to the real performance of individual models. For this reason, we consider the difference in each criterion for each individual data set for both the model under consideration together with the MCMC analysis of $\Lambda$CDM.

\subsection{Observational Constraints}

In this section, we consider how the quadratic Palatini model fairs with observational data through an MCMC constraint analysis. We do this for a combination of all the data sets described previously with CC being our baseline for all analyses. We also perform these analyses again with the $H_0^{\rm S22}$ prior on the Hubble constant, as well as including statistical criterions in order to compare and contrast the different performances.

\begin{figure}[h]
    \centering
    \includegraphics[width = 0.75\textwidth]{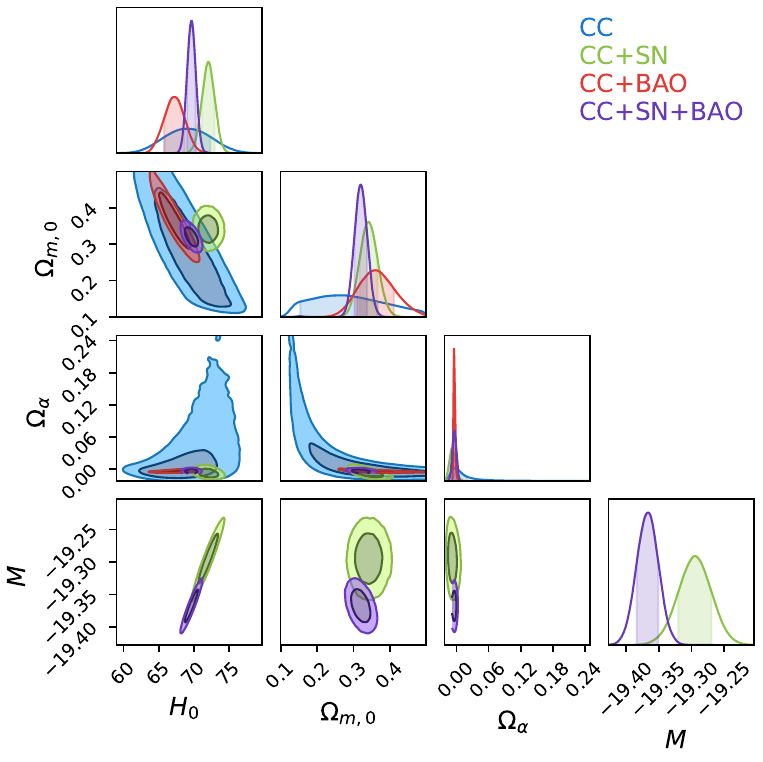}
    \caption{Posteriors and confidence contours for the model parameters for a combination of data sets. The blue and green contours show the CC with Pantheon+ and CC with BAO data sets respectively, while the red contour shows the full combination.}
    \label{fig:Model}
\end{figure}

Considering the Friedmann equation in Eq.~\eqref{friedmann2}, we can fit for the set of free parameters that make up this model. We assume a flat FLRW geometry to avoid additional degeneracies. By evaluating this relation at current times, we find that 
\begin{align}
    &\Omega_{\Lambda} = \Big[-1-4\Omega_{\alpha}(-2 + \Omega_{m,0}) \pm \nonumber\\
    &\sqrt{1 + \Omega_{\alpha}^2 \left(64 - 96 \Omega_{m,0}\right) - 8\Omega_{\alpha} \left(-2 + 3\Omega_{m,0} + 4\Omega_{r,0}\right)}\Big]/16\Omega_{\alpha}\,,
\end{align}
which has a healthy $\Lambda$CDM as $\Omega_{\alpha}\rightarrow 0$. The analytical nature of the solution in Eq.~\eqref{friedmann2} makes it very fast, and more able to be incorporated into precision calculations. We show the constraints for the full set of parameters, namely $\{H_0, \Omega_{m,0}, M, \Omega_{\alpha}\}$ in Fig.~\ref{fig:Model}. The high level of consistency in the posterior of the $\Omega_{\alpha}$ shows promising results for the low-energy Palatini model. On the other hand, the parameter seems to be degenerate with the other parameters to a certain degree, except for the CC+SN+BAO case where the contours are rather small. For the other parameters, similar relationships are observed as in the standard cosmological model case. For instance, the nuisance parameter is pushed to lower values when BAO data is incorporated, while an anti-correlation is observed between the Hubble constant and the matter density parameter. The matter density parameter is also very consistent across the different data sets, while the posteriors of the Hubble constant show a mild tension between the joint data sets.

The outputs of the MCMC constraint analysis are shown in Table~\ref{tab:model_outputs}, while the corresponding $\Lambda$CDM fits are contained in Appendix~\ref{App:LCDM_constraints}. One notes that higher Hubble constants are obtained in the quadratic Palatini cases except when BAO data is included. On the other hand, smaller matter density parameters are obtained except for the baseline case. As for the $\Omega_{\alpha}$ parameter, the observational data appears to prefer values that are in the $1-2\sigma$ range as compared with the $\Lambda$CDM value where this parameter is zero. Indeed, for the case where the statistical distance is largest from $\Lambda$CDM, namely for the CC+BAO data set, the statistical criterions give a negative value indicating that the model is slightly preferred, whereas the other cases show a marginal preference for the standard cosmological model.

\begin{table*}
    \centering
    \caption{Results for the model where the first column lists the data set combinations, the second column contains the Hubble constant constraint, while the third, fourth, and fifth columns contain the matter density, nuisance, and $\Omega_{\alpha}$ parameters. The sixth and seventh columns show the difference in AIC and BIC criterions as compared with the corresponding $\Lambda$CDM constraints.}
    \label{tab:model_outputs}
    \begin{tabular}{ccccccc}
        \hline
		Data sets & $H_0$ & $\Omega_{m,0}$ & $M$ & $\Omega_{\alpha}$ & $\Delta {\rm AIC}$ & $\Delta {\rm BIC}$ \\ 
		\hline
        CC & $68.9^{+3.4}_{-3.1}$ & $0.267^{+0.099}_{-0.113}$ & -- & $-0.0030^{+0.0230}_{-0.0120}$ & 1.97 & 1.46\\ 
		CC+SN & $72.06^{+0.86}_{-0.98}$ & $0.341^{+0.024}_{-0.025}$ & $-19.295^{+0.026}_{-0.025}$ & $-0.0084^{+0.0060}_{-0.0049}$ & 0.08 & 1.32 \\ 
		CC+BAO & $67.3^{+1.4}_{-1.5}$ & $0.361\pm 0.051$ & -- & $-0.0045^{+0.0016}_{-0.0012}$ & $-2.84$ & $-3.14$ \\ 
		CC+SN+BAO & $69.64^{+0.60}_{-0.64}$ & $0.320^{+0.017}_{-0.018}$ & $-19.366^{+0.015}_{-0.019}$ & $-0.0042\pm 0.0040$ & 2.38 & 3.62 \\ 
		\hline
    \end{tabular}
\end{table*}

\begin{figure}[h]
    \centering
    \includegraphics[width = 0.75\textwidth]{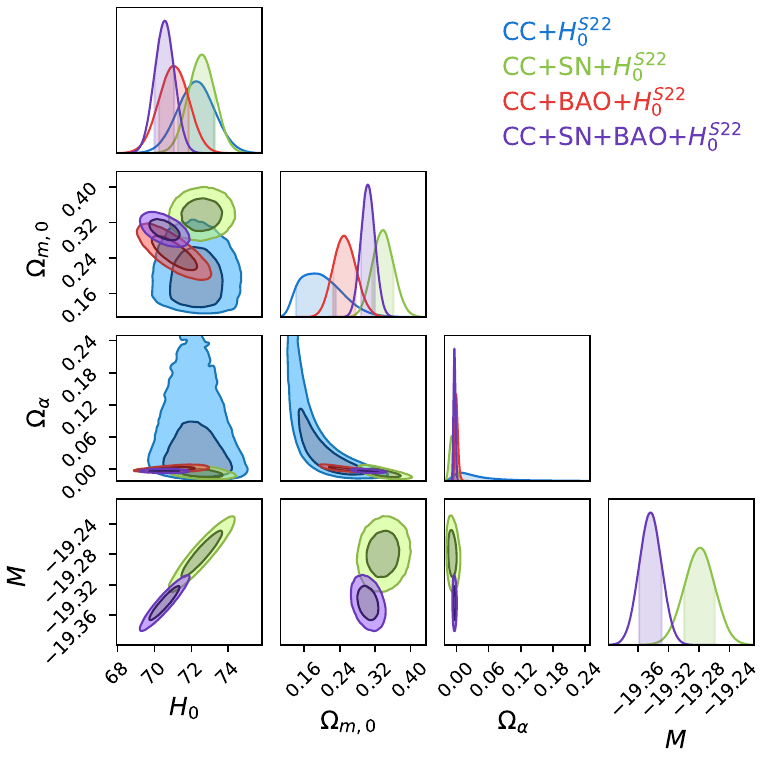}
    \caption{The posteriors and confidence contours for the model parameters are again shown except that the $H_0^{\rm S22}$ prior added in the constraint analysis.}
    \label{fig:Model_S22}
\end{figure}

We consider the case of adding the $H_0^{\rm S22}$ prior to all the data sets under consideration in Fig.~\ref{fig:Model_S22} to assess the consistency of the analyses with this value of the Hubble constant reported in the literature. The introduction of the prior makes the absolute magnitude and Hubble constant anti-correlation more pronuanced while the consistency of the matter density parameter is slightly perturbed. On the other hand, the posterior values of the $\Omega_{\alpha}$ parameter are generically more consistent with $\Lambda$CDM in this scenario.

\begin{figure}[h]
    \centering
    \includegraphics[width = 0.75\textwidth]{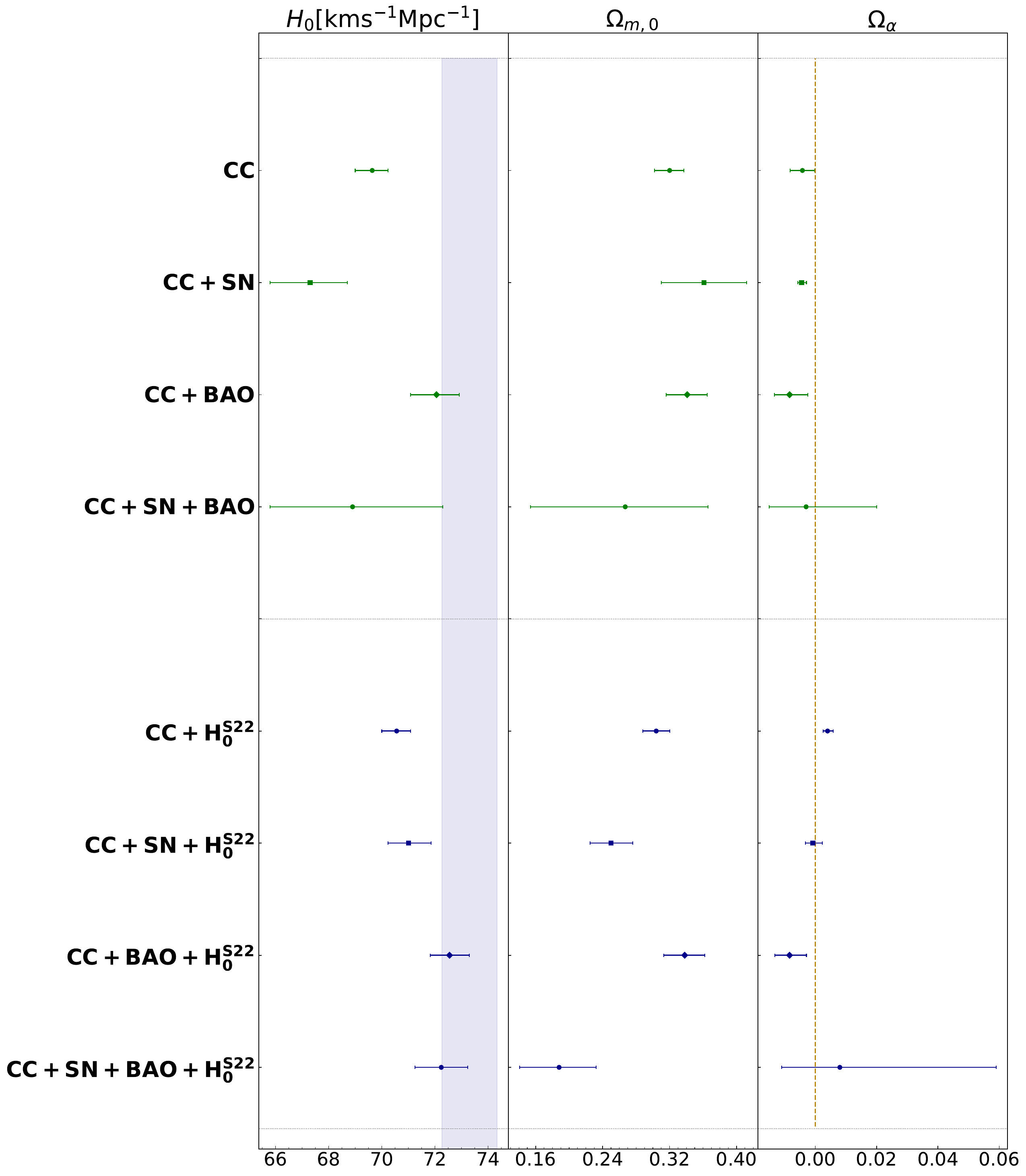}
    \caption{Comparative constraints for the different data sets under consideration together with the corresponding outputs when the $H_0^{\rm S22}$ prior is included. The shaded region shows the 1$\sigma$ uncertainty of the $H_0^{\rm S22}$ Hubble constant value.}
    \label{fig:summary}
\end{figure}

The outputs of the MCMC analyses of the Palatini model for the $H_0^{\rm S22}$ prior case are shown in Table~\ref{tab:model_outputs_S22}. Generically, we find higher values of the fits for the Hubble constant, as one would expect, but we also find lower values of the matter density parameters with reasonable uncertainties, which is consistent with how one would expect the cosmology to evolve. On the other hand, the precise fits for the $\Omega_{\alpha}$ parameter are consistent with $\Lambda$CDM but there is a preference for a nontrivial $\Omega_{\alpha}$. By and large the statistical performance of the model against the different data sets remains largely the same. The superior performance of the model for the instance where we consider CC+BAO data set is considered remains the case. The various fits for the model parameters and the different data set combinations are shown in Fig.~\ref{fig:summary} which contains a whisker plot of the final constraints together with an indicative region for the $H_0^{\rm S22}$ prior.

\begin{table*}
    \centering
    \caption{The corresponding results for the model when the $H_0^{\rm S22}$ prior is included in the constraint analysis.}
    \label{tab:model_outputs_S22}
    \begin{tabular}{ccccccc}
        \hline
		Data sets & $H_0$ & $\Omega_{m,0}$ & $M$ & $\Omega_{\alpha}$ & $\Delta {\rm AIC}$ & $\Delta {\rm BIC}$ \\ 
		\hline
        CC & $72.24^{+1.00}_{-0.99}$ & $0.188^{+0.044}_{-0.047}$ & -- & $0.0080^{+0.0510}_{-0.0190}$ & 1.32 & 0.81 \\ 
		CC+SN & $72.55^{+0.74}_{-0.72}$ & $0.338^{+0.024}_{-0.025}$ & $-19.278^{+0.019}_{-0.021}$ & $-0.0084^{+0.0055}_{-0.0048}$ & 0.01 & 1.25 \\ 
		CC+BAO & $71.01^{+0.85}_{-0.77}$ & $0.250^{+0.026}_{-0.025}$ & -- & $-0.0008^{+0.0031}_{-0.0024}$ & -1.94 & -2.23 \\ 
		CC+SN+BAO & $70.56^{+0.52}_{-0.56}$ & $0.304\pm 0.016$ & $-19.343^{+0.014}_{-0.015}$ & $0.0040^{+0.0018}_{-0.0015}$ & 7.05 & 8.29 \\ 
		\hline
    \end{tabular}
\end{table*}

\section{Conclusion}

The leading order corrections to GR in the Palatini approach scenario have been probed in this work using the latest expansion rate data with an emphasis on the latest SNIa and BAO data. The Palatini formulation of gravity offers an interesting approach in which the relationship between the gravitational connection and the metric tensor field remains open and may be used as a freedom of the system to address more physical constraints. This is important so that the various implicit assumptions in the standard formulation of gravity can be isolated and explored individually. In our case, we explore the late Universe and so we can assume the low-energy limit of Palatini gravity, which produces the Friedmann equations as explained in Sec.~\ref{sec:palatini_cosmo} where the gravitational sector leads directly to the appearance of an additional parameter $\Omega_{\alpha}$ which is responsible for the low-energy gravitational action corrections. For our analyses, we use CC data as our baseline data set with contributions also from the SN sample and a combination of transversal BAO data sets, that are non-overlapping. For each combination of data sets, we perform a full MCMC analysis obtaining constraints on all cosmological parameters of the model under investigation. Additionally, we use the latest literature value of the Hubble constant from the SH0ES Team, namely $H_0^{\rm S22}$. Finally, we give a statistical comparison of the Palatini quadratic model against $\Lambda$CDM using the AIC and BIC metrics. 

The $\Lambda$CDM reference constraint values are contained in Appendix~\ref{App:LCDM_constraints}, which is important for estimating the AIC and BIC statistical measures for the different data set combinations. In general, our analysis shows promising results for this low-energy limit of Palatini gravity. We find higher values of the Hubble constant and consistent matter density parameter ranges. As for the new $\Omega_{\alpha}$, this is mostly consistent with the $\Lambda$CDM limit but also shows differences in some cases, which may be further nuanced with the addition of data sets. Considering the AIC and BIC measures, these seem to show a mild preference for $\Lambda$CDM in most cases since it has less model parameters in this setting. However, for the CC+BAO data set combination, we find a slight preference for the Palatini quadratic model which motivates us to further study this model. These features are strengthened when the $H_0^{\rm S22}$ prior is added to the analysis being undertaken.

Let us compare our findings to the results obtained previously in the literature. In the SI units, the theory parameter $\alpha$ can be expressed as $\alpha = \frac{c^2 \Omega_\alpha}{3H^2_0}$; for example, for $\Omega_\alpha \sim 10^{-3}$ we get $\alpha \sim 10^{48}\, \text{m}^2$. If we consider the case where a slight preference for the quadratic model is exhibited, i.e. for the CC+BAO data set, we get the following range for the $\alpha$ parameter: $\alpha = -0.496^{+1.921}_{-1.488}\times 10^{49}\,\text{m}^2$. This gives a better bound on the parameter related to the quadratic correction than found in \cite{Pinto:2018rfg} using Union2.1 data set: $\alpha = -0.092^{+8.840}_{-9.025}\times 10^{55}\,\text{m}^2$, although the model considered there features also as additional term proportional to $1/\mathcal{R}$. The same model as ours was analyzed also in \cite{Stachowski:2016zio}; the data used comprised of CMB observations, data from the WMAP, 580 SNIa events, and measurements of $H(z)$; the results were $\Omega_\alpha = 9.70^{+134.80}_{-9.70}\times 10^{-9}$. For this model, the BIC value was greater than 6, indicating a preference for the $\Lambda$CDM. In \cite{Borowiec:2011wd}, utilizing observational $H(z)$, BAO, and CMB data, the Lagrangian of the form $L = \sqrt{-g}(\mathcal{R} + \alpha \mathcal{R}^2 + \beta \mathcal{R}^{1 + \delta} + \gamma \mathcal{R}^{1 + \sigma} L_m)$ was considered; it was found that, depending on the assumptions, the dimensionless parameter varies from $\Omega_\alpha = -56.342^{+3.102}_{-2.971}$ to $\Omega_\alpha = 4.401\pm 0.079$. In \cite{Borowiec:2015qrp}, a similar methodology was used to assess the goodness of the Starobinsky model with a modified EoS - instead of the perfect fluid, the so-called Chaplygin gas was used, mimicking the effects of dark matter at high densities, and dark energy at the low end of the energy scale; the parameter range is $\Omega_\alpha = -1.156^{+1.156}_{-0.08} \times 10^{-9}$; the model, however, features three new parameters, as compared to just one introduced by the modification of the Einstein-Hilbert Lagrangian. {Summarizing, the latest cosmological data allows us to put the bound on the parameter $\alpha$ in the range  $|\alpha| \leq 10^{49}$ m$^2$ }.

On the other hand, in the context of the $f(\mathcal{R})=\mathcal{R}+\alpha \mathcal{R}^2$ model, investigations into the weak-field limit have indicated that the magnitude of $\alpha$ typically remains below $2\times 10^{8}\, \text{m}^2$ \cite{Olmo:2005zr}. Nevertheless, precise constraints on these parameters have proven challenging due to uncertainties stemming from microphysics, rendering experiments within the Solar System inadequate for such determinations \cite{Toniato:2019rrd}. It is important to note that gravitational tests conducted in a vacuum, such as the Shapiro delay technique, do not impose constraints on Palatini gravity since the theory reduces to General Relativity with a cosmological constant, as elucidated in the discussion following Eq. \eqref{struc}.

In a manner reminiscent of General Relativity, none of the $f(\mathcal{R})$ models effectively account for the rotation curves observed in galaxies \cite{Hernandez-Arboleda:2022rim,Hernandez-Arboleda:2023abv}. Consequently, constraints derived from galaxy catalogs continue to remain elusive up to the present day.

Conversely, when delving into microphysical considerations, seismic data from Earth has imposed more stringent limitations on the parameter, with $|\alpha|\lesssim 10^9 \,\text{m}^2$ (at a $2\sigma$ level of precision) \cite{Kozak:2023axy,Kozak:2023ruu}. Notably, in studies focused on the non-relativistic limit, it has been ascertained that only the quadratic term holds significance, while higher-order terms, starting from the sixth order, become relevant \cite{Toniato:2019rrd}. Therefore, these bounds also hold for the model considered here. 

The significant divergence of nearly 40 orders of magnitude in the constraints obtained underscores the imperative need for an intensified exploration of modified gravity within the context of stellar and substellar objects. This demand is further underscored by the multifaceted complexities inherent in astrophysical modeling, necessitating a realistic treatment of matter descriptions that account for potential anisotropies, non-ideal fluid characteristics, energy transport mechanisms, opacity considerations, and more. Furthermore, it is essential to address uncertainties related to the material component, a factor often disregarded in cosmological contexts.

In accordance with the insights presented in \cite{Baker:2014zba}, particularly evident in diagrams 1 and 2, the curvature regime governing these celestial bodies occupies a pivotal intermediate position. This region bridges the divide between the problematic cosmological realm and the domain where GR suffices to describe gravitational phenomena adequately. Consequently, this transitional realm, serving as a boundary between small-scale systems like compact objects and the Solar System, may conceal the subtle emergence of corrections to GR.

Nevertheless, the low-energy Palatini model analysis being showcased here gives precision insights into the behavior of the model in the context of the late Universe together with its expansion profile in that regime. The study gives a positive sign that this approach shows promise in confronting the major challenges that have appeared in the literature in recent years. We intend to further advance in this direction with a broader numerical analysis of this model including perturbations and related data sets in the near future.

\appendix

\section{\texorpdfstring{$\Lambda$}{}CDM Constraints}
\label{App:LCDM_constraints}
The parameter constraints for the $\Lambda$CDM baseline model are shown in Table~\ref{tab:LCDM_outputs}. These standard cosmological model constraints can be compared with the constraints obtained in the main sections of this work in order to understand how the cosmic evolution changes in this scenario.

\begin{table*}
    \centering
    \caption{Parameter constraints on $\Lambda$CDM for different combinations of data sets.}
    \label{tab:LCDM_outputs}
    \begin{tabular}{cccccc}
        \hline
		Data sets & $H_0$ & $\Omega_{m,0}$ & $M$ & AIC & BIC \\ 
		\hline
        CC & $67.7^{+3.0}_{-2.9}$ & $0.321^{+0.062}_{-0.058}$ & -- & 20.523 & 18.997 \\ 
		CC+SN & $71.68^{+0.84}_{-0.93}$ & $0.320^{+0.015}_{-0.017}$ & $-19.301^{+0.026}_{-0.024}$ &  1561.118 & 1564.833 \\ 
		CC+BAO & $68.74^{+0.79}_{-0.80}$ & $0.294\pm 0.012$ & -- & 34.703 & 33.826 \\ 
		CC+SN+BAO & $70.57\pm 0.64$ & $0.310\pm 0.010$ & $-19.336\pm 0.020$ & 1583.535 &  1587.266 \\ 
        \hline
		CC+$H_0^{\rm S22}$ & $67.7^{+3.0}_{-2.9}$ & $0.321^{+0.062}_{-0.058}$ & -- &    23.163902 &    21.637987 \\ 
		CC+SN+$H_0^{\rm S22}$ & $71.68^{+0.84}_{-0.93}$ & $0.320^{+0.015}_{-0.017}$ & $-19.301^{+0.026}_{-0.024}$ &  1562.141016 &  1565.856660 \\ 
		CC+BAO+$H_0^{\rm S22}$ & $68.74^{+0.79}_{-0.80}$ & $0.294\pm 0.012$ & -- &    45.391895 &    44.514606 \\ 
		CC+SN+BAO+$H_0^{\rm S22}$ & $70.57\pm 0.64$ & $0.310\pm 0.010$ & $-19.336\pm 0.020$ &  1587.608045 &  1591.338647 \\ 
  \hline
    \end{tabular}
\end{table*}

\section*{Acknowledgements}
This article is based upon work from COST Action CA21136 Addressing observational tensions in cosmology with systematics and fundamental physics (CosmoVerse) supported by COST (European Cooperation in Science and Technology).
 AK and DAG were beneficiaries of Short Term Scientific Missions (STSM) of the COST action CosmoVerse (CA21136). JLS would like to acknowledge funding from ``The Malta Council for Science and Technology'' through the ``FUSION R\&I: Research Excellence Programme''. AW acknowledges financial support from MICINN (Spain) {\it Ayuda Juan de la Cierva - incorporaci\'on} 2020 No. IJC2020-044751-I.

\end{document}